\documentclass{sf2a-conf2017}
\usepackage[utf8]{inputenc}
\usepackage{graphicx}
\usepackage{hyperref}
\usepackage[]{natbib}  
\usepackage{amsmath}
\usepackage{enumitem}
\usepackage{physics}
\usepackage{amssymb}

\def\BibTeX{{\rm B\kern-.05em{\sc i\kern-.025em b}\kern-.08em
    T\kern-.1667em\lower.7ex\hbox{E}\kern-.125emX}}
\bibpunct{(}{)}{;}{a}{}{,}  %%%%%%%%%%%%%  A&A bibliography style

\newcommand{\ft}{\widetilde{f}}
\newcommand{\mA}{\mathcal{A}}
\newcommand{\mB}{\mathcal{B}}

\newcommand{\sslash}{\mathbin{\!/\mkern-5mu/\!}}
\newcommand{\kp}{k_{\sslash}}
\newcommand{\dU}{\frac{\dd\overline{U}}{\dd y}}
\newcommand{\Vc}{\hat{V}}
\newcommand{\ddl}{\frac{\dd\delta}{\dd y}}
\renewcommand{\sc}{\hat{\sigma}}
\newcommand{\kx}{k_{x}}
\newcommand{\ddU}{\frac{\dd^2\overline{U}}{\dd y^2}}
\newcommand{\Um}{\overline{U}}
\newcommand{\yc}{y_{\mathrm{c}}}
\DeclareMathOperator{\e}{e}
\newcommand{\Ek}{E_\mathrm{k}}
\newcommand{\mK}{\mathcal{K}_\mathrm{c}}
\newcommand{\I}{\mathrm{I}}

\newcommand{\It}{\mathrm{III}}
\renewcommand{\sf}{\sigma_\mathrm{f}}
\begin{document}

\TitreGlobal{SF2A 2017}

%%-----------------------------------------------------------------
\title{Towards a better understanding of tidal dissipation at corotation layers in differentially rotating stars and planets}

\runningtitle{Tidal dissipation at corotation layers}

\author{A. Astoul}\address{Laboratoire AIM Paris-Saclay, CEA/DRF - CNRS - Université Paris Diderot, IRFU/DAp Centre de Saclay, F-91191 Gif-sur-Yvette, France}
\author{S. Mathis$^1$}
\author{C. Baruteau}\address{IRAP, Université de Toulouse, CNRS, UPS, Toulouse, France}
\author{Q. André$^1$}

%% Keep this line, even if the page will be settled afterwards.
\setcounter{page}{237}

%%-----------------------------------------------------------------

\maketitle

%%-----------------------------------------------------------------
%%        The abstract
%% 
\begin{abstract}Star-planet tidal interactions play a significant role in the dynamical evolution of close-in planetary systems. We investigate the propagation and dissipation of tidal inertial waves in a stellar/planetary convective region. We take into account a latitudinal differential rotation for the background flow, similar to what is observed in the envelope of low-mass stars like the Sun. Previous works have shown that differential rotation significantly alters the propagation and dissipation properties of inertial waves. In particular, when the Doppler-shifted tidal frequency vanishes in the fluid, a critical layer forms where tidal dissipation can be greatly enhanced. Our present work develops a local analytic model to better understand the propagation and dissipation properties of tidally forced inertial waves at critical layers.
\end{abstract}

\begin{keywords}
Hydrodynamics -- Waves -- Planet-star interactions -- Planets and satellites : dynamical evolution and stability -- Stars : rotation
\end{keywords}

%%-----------------------------------------------------------------

\section{Introduction}
The study of tides and especially of tidal dissipation is essential to understand the secular evolution of planetary systems. Inertial waves, which are tidally excited in convective zones, carry and deposit energy and angular momentum in the envelope of low-mass stars. This dissipation can be modified by  the internal dynamics of the convective envelope like differential rotation as pointed out by \citet{BaruteauRieutord} and \citet{Guenel}. In a uniformly rotating fluid body, \citet{Favier} have also shown that differential rotation arises from non-linear evolution of inertial waves excited by tides. In the solar convective envelope, the rotation rate depends mostly on the colatitude with a difference in rotation rate of $\sim30\%$ between the equator and the pole \citep[e.g.][]{Garcia}. \citet{Guenel} and \citet{GuenelarXiv}, following the work of \citet{BaruteauRieutord}, examined the viscous dissipation of tidally excited inertial waves for conical solar and anti-solar rotation profiles. They found that differential rotation modifies the characteristics and amplitude of the tidal dissipation particularly at critical layers.
Nevertheless, a better physical understanding is required to explain the influence of these layers and the different regimes found in numerical simulations. To reach this objective we develop a new local Cartesian model describing a small fraction of the differentially rotating convective envelope in a low-mass star. This model allows us to focus on small-scale effects, to understand how latitudinal sheared flow impacts the propagation of tidal inertial waves, as well as the role of critical layers.

\section{Local Cartesian model for tidal waves propagating in a convective fluid with latitudinal shear}
\subsection{Presentation of the model and main assumptions}
\begin{figure}[t!]
\centering
\includegraphics[scale=0.6]{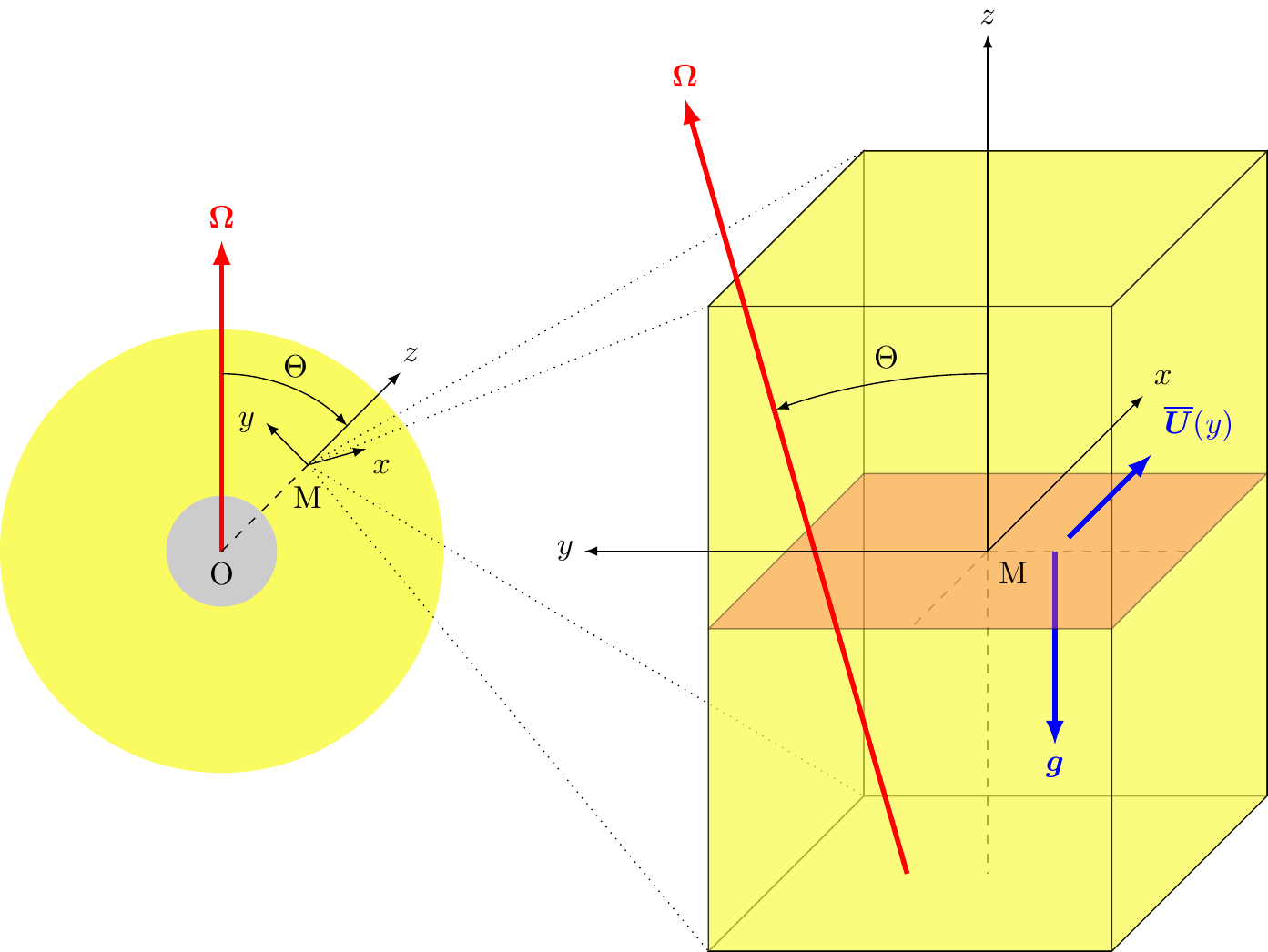}
\caption{Left : global view of a solar-like star. The convective envelope, depicted in yellow, lies on top of a stably stratified core (grey area). Right : the local Cartesian box, centered on a point M in the envelope, corresponding to a colatitude $\Theta$. The local box is tilted with respect to the spin axis. Its vertical axis $z$, corresponding to the local radial direction, is anti-aligned with the gravity $\bf g$. The $x$ and $y$ axes correspond to the local azimuthal and latitudinal directions, respectively. [Adapted from \citet*{Andre}]}
\label{astoul:fig1}
\end{figure}

The concept of a local box to study tidal waves has been introduced by \citet{AuclairDesrotour} following the work of \citet{OgilvieLin}. The box is described with a set of local Cartesian coordinates $\{x,y,z\}$ centered on a specific point M of the convective envelope, and is inclined with respect to the star's rotation axis (as illustrated in Fig. \ref{astoul:fig1}). The dimensions of our box are taken to be small with respect to the characteristic length scale of the convective envelope, in order to remove curvature effects. The latitudinal shear is embodied by an azimuthal mean flow velocity, ${\bf \overline U}=\overline U(y){\bf e}_x$.
In addition, the uniform rotation vector $\boldsymbol{\Omega}$ makes an angle $\Theta$ with our local vertical axis so that we have $2\boldsymbol{\Omega}=(0,2\Omega\sin\Theta,2\Omega\cos\Theta)=(0,\ft,f)$. 
Moreover, we make the following assumptions :
 \begin{itemize}[label=$\ast$]
\item we adopt the Boussinesq approximation. Hence, the continuity equation simplifies to ${\div \bf u} =0$, where $\bf u$ is the local velocity of the fluid.
\item we assume that in convective layers, the dissipation is driven by an effective viscous-like turbulent friction $\nu_\mathrm{eff}$ that leads us to neglect as a first step thermal diffusion \citep[we refer the reader to Fig. 9 in][for a detailed discussion]{AuclairDesrotour}. Furthermore, we simplify the viscous diffusion term $\nu_\mathrm{eff} \Delta \bf u$ as a linear (Rayleigh) friction term, which we write as $\sf \bf u$ \citep{Ogilvie2009} with $\sf$ homogeneous to a pulsation.
\item the centrifugal acceleration is neglected. This is acceptable for not to fast rotators.
\item we carry out a linear analysis, and therefore all quantities associated with the base flow remain unaltered. \end{itemize}
\subsection{System of equations}
At first order, all dynamical quantities can be split into two components, an unperturbed background quantity and a small perturbation. As we study inertial waves we are interested in the dynamical tide. The equations for the perturbations of the hydrostatic balance by the companion are thus treated elsewhere \citep[see e.g.][]{Ogilvie2013}. All perturbed quantities are assumed to be periodic in time and along the local azimuthal axis. As a result, projected on the Cartesian basis, the Navier-Stokes and continuity equations in the rotating frame write :
\begin{align}
\left\{\begin{aligned}
\Sigma u + \left(\frac{\partial\overline U}{\partial y} - f\right)v +\widetilde{f}w & =  - i k_x \Pi +F_x,\\ 
\Sigma v +fu & = -\frac{\partial\Pi}{\partial y} +F_y,\\
\Sigma w -\widetilde{f}u & = -\frac{\partial\Pi}{\partial z}  +F_z, \\
ik_xu +\partial_y v+\partial_z w & = 0,
\end{aligned}\right.
\label{set}
\end{align}
where we define the quantities : ${\bf u}=(u,v,w)$ and $\Pi=p/\overline{\rho}$, which stand for the perturbed velocity of the fluid in the box and the local perturbed pressure divided by the mean density, respectively. Moreover, we denote by ${\bf F}=(F_x,F_y,F_z)$ the tidal acceleration which operates throughout the fluid. Finally, we have introduced a complex frequency $\Sigma = i\hat{\sigma}+\sigma_f$, with $\hat{\sigma}=\omega+k_x\overline{U}(y)$ the Doppler-shifted frequency,  $\omega$ the frequency of the excited waves in the inertial frame and $k_x$ the azimuthal wave number. 
\subsection{The Poincaré equation}
We solve our set of Equations (\ref{set}) by the substitution method, only keeping the second-order derivatives (i.e. the WKBJ approximation).
Consequently, we derive from the system (\ref{set}) the so-called Poincaré equation for tidally forced inertial waves propagating in a latitudinal shear : 
\begin{equation}\label{pointc}
\begin{split}
\underbrace{\left(\Sigma^2+\widetilde{f}^2\right)}_{\mathcal{A}}\partial_{y,y}v+\underbrace{\widetilde{f}\left(2f-\partial_y\overline{U}\right)}_{\mathcal{B}}\partial_{y,z}v +\underbrace{\left[\Sigma^2+f(f-\partial_y\overline{U})\right]}_{\mathcal{C}}\partial_{z,z}v=S(x,y,z,t),
\end{split}
\end{equation}
with $S(x,y,z,t)$ the complex source term due to the tidal forcing, which does not depend on the velocity components.

For $\mA=\Sigma^2+\widetilde{f}^2\neq0$, we introduce the transformation : $v(x,y,z,t) = \hat{V}(y)\exp{i(\kp z+\delta(y)+\kx x+\omega t)}$, following \citet{GerkemaShriraJGR}, where $\delta$ satisfies $\ddl=-\kp\frac{\mB}{2\mA}$.
This allows us to simplify the Poincaré equation (\ref{pointc}) as :
\begin{equation}
\frac{\dd^2\Vc}{\dd y^2}+\underbrace{\frac{\kp}{4\mA^2}\left[2i\ft\ddU\mA-4\Sigma\kx\dU\mB+\kp(\mB^2-4\mathcal{AC})\right]}_{\kappa(y)^2}\Vc=\frac{\hat S(y)}{\mA},
\label{Schrofinal}
\end{equation}
where $S(x,y,z,t)$ has been projected on a basis of orthogonal functions \citep[see][]{GerkemaShriraJFM,MathisNeiner}.
From a physical point of view, this second-order ordinary differential equation is similar to a Schrödinger equation with a complex potential $\kappa(y)$.

\section{The key role of critical layers}
We now propose to investigate the solutions to the modified Poincaré equation (\ref{Schrofinal}) near the points where $\mA$ cancels out, to understand waves propagation in the vicinity of such critical layers.
\subsection{The modified Poincaré equation}
In the non-rotating adiabatic case, a singularity is found for $\sc=0$ \citep{BookerBretherton}. For a vertical or inclined rotation vector, \citet{Jones} and \citet{Grimshaw} have shown that a singularity is obtained for $\sc=-2\Omega$, $0$ and $2\Omega$. Because of differential rotation and viscosity, our critical layers are different. For a vanishing viscosity, the singularities are (i) for $\sc^2=\ft^2$, as can be noticed by setting $\mA=0$ in the above Poincaré equation, and (ii) for $\sc=0$, which can be seen from the polarization relationships (not written here).
To expand the Poincaré equation (\ref{Schrofinal}) near the critical layers $y=\yc$ corresponding to $\mA=0$, we adopt the method used by \citet{Alvan} in the case of internal gravity waves propagating in a fluid with a vertical shear.
Taking the Taylor expansion of $\mA$ in the vicinity of $\yc$ at first order, we can express the Poincaré equation (\ref{Schrofinal}) for free inertial waves around critical layers as follows :
\begin{equation}
\Vc''(y)+\frac{\chi}{(y-\yc)^2}\Vc(y)=0,
\label{Schro3}
\end{equation}
where the complex variable $\chi$ gathers constant quantities evaluated at $\yc$. It mainly depends on the Rossby number\footnote{which evaluates the competition between the shear and the Coriolis acceleration.} of the differential rotation $Ro = \Um'(\yc)/(2\Omega)$, and the Ekman number\footnote{It describes the relative strength of viscous forces and the Coriolis acceleration.} $E_k=\sf/(2\Omega)$.

\subsection{Applying the method by \citet{Alvan}}
 Let us write $\Vc(y)=(y-\yc)^r$, where $r$ is an unknown complex number. Injecting this function in Eq. (\ref{Schro3}), we obtain the index equation : $r(r-1)+\chi=0$.
If we consider the case where $\chi$ is real, two cases are possible depending on the sign of the discriminant $\Delta=1-4\chi$. The ensuing criterion is similar to the one found by \citet{Alvan} involving the Richardson number.

\subsubsection{The stable regime}
We first look at the case where the discriminant $\Delta$ is negative (i.e. $\chi>1/4$).
The solution in the vicinity of $\yc$ is :
\[\Vc(y)=\alpha(y-\yc)^{\frac{1}{2}+i\sqrt{\chi-\frac{1}{4}}}+\beta(y-\yc)^{\frac{1}{2}-i\sqrt{\chi-\frac{1}{4}}},
\]
with $\alpha$ and $\beta$ the amplitude of the wave function that can be seen as a combination of upward and downward-propagating waves. In order to know the behaviour of the wave getting across the critical layer, we can reconnect the solution below and above this layer.
A wave propagating upward through the critical layer, is attenuated by a factor $\exp{-\pi \sqrt{\chi-\frac{1}{4}}}$ and dephased by an argument $\pi/2$. Likewise a wave propagating in the opposite direction, is attenuated by the same factor and has a phase difference of $-\pi/2$. As a consequence, we identify a stable regime. We plot this coefficient of attenuation in the upper panel of Fig. \ref{astoul:fig234} for different colatitudes. We observe that the attenuation is greater for low Rossby numbers which correspond to fast rotating stars for a given shear, or to weak differential rotation at fixed global rotation. Moreover, at fixed Rossby number,  the attenuation is larger for a weak colatitude, that is near the rotation axis.

\subsubsection{The unstable regime}
\begin{figure}[ht!]
\centering
\includegraphics[width=0.48\textwidth,clip]{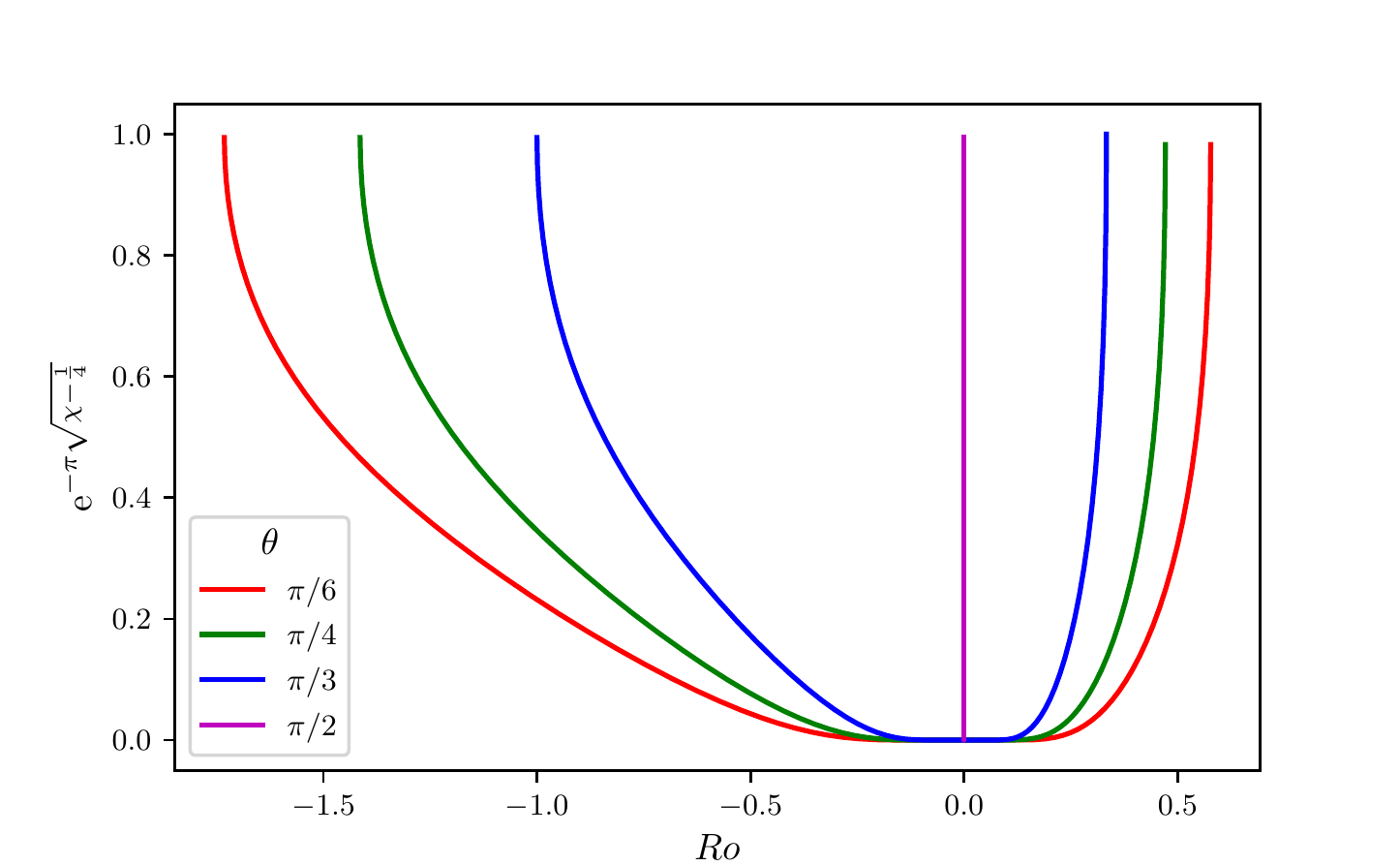}\\
\includegraphics[width=0.48\textwidth,clip]{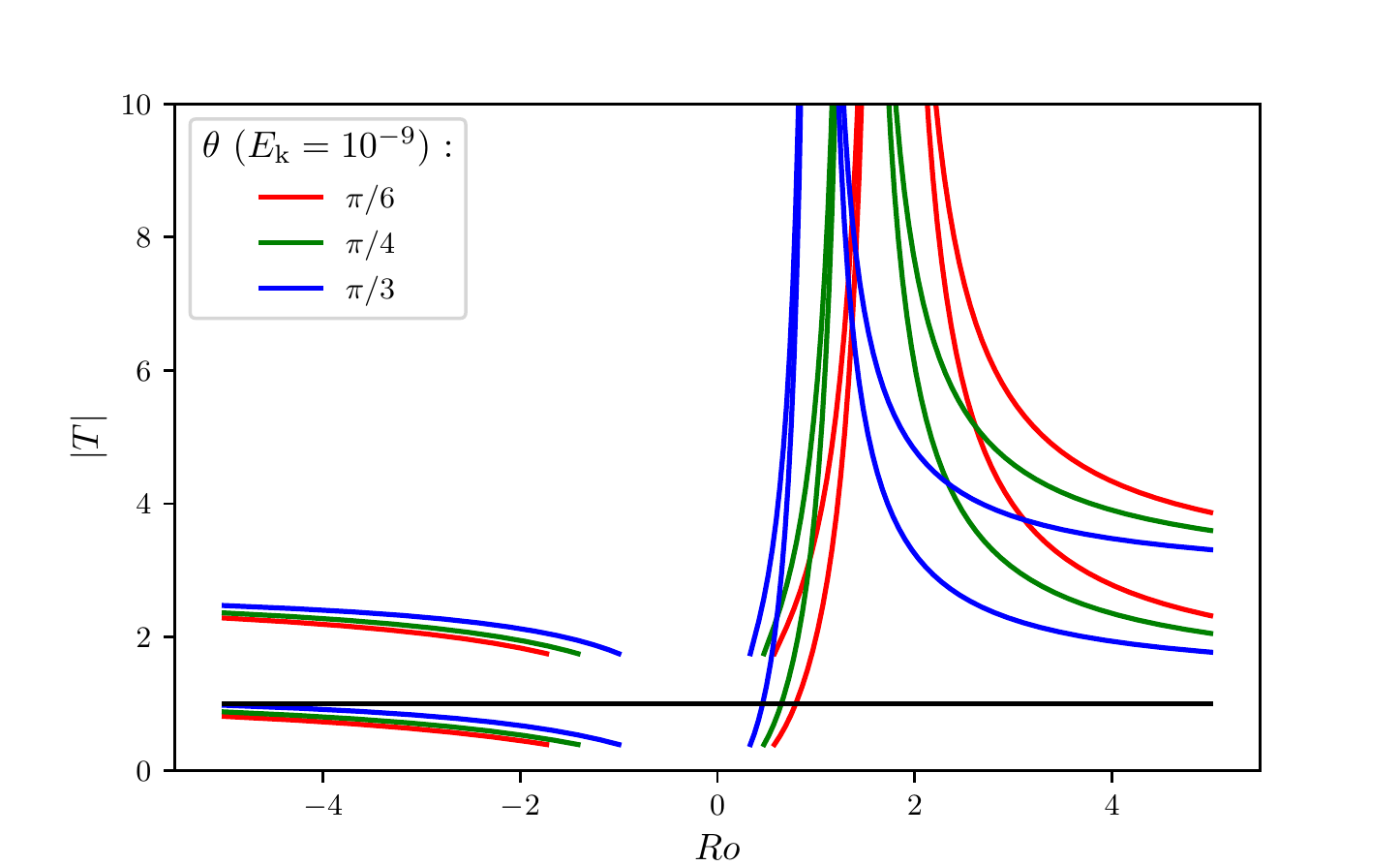}
\includegraphics[width=0.48\textwidth,clip]{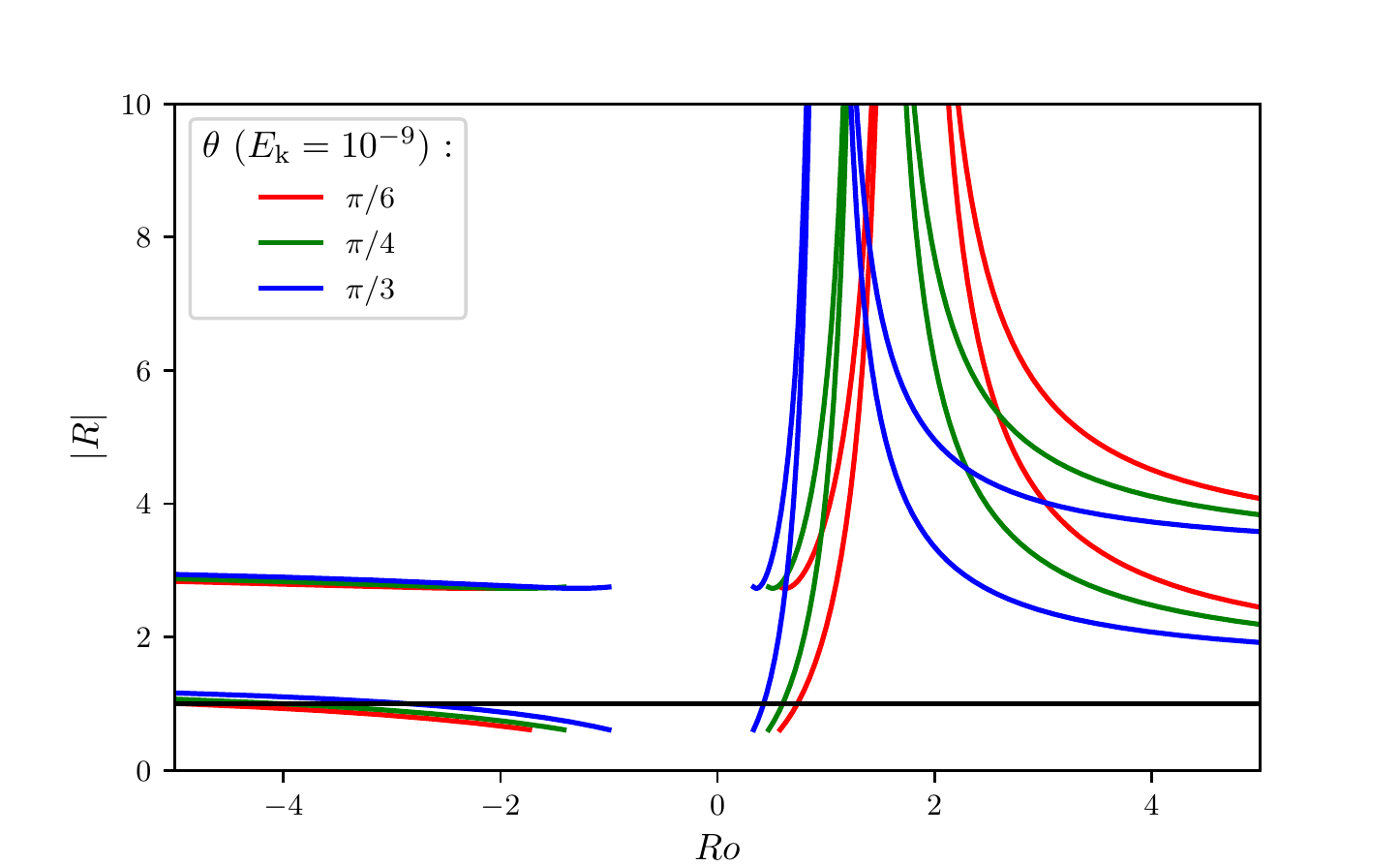}
\caption{{\bf Top :} Attenuation coefficient as a function of the Rossby number for different values of the colatitude $\Theta$. The Ekman number is set to $E_k=10^{-9}$.
{\bf Bottom :} Absolute value of the transmission (left) and reflexion (right) coefficients as a function of the Rossby number. These coefficients are displayed for different colatitudes and for $\Ek=10^{-9}$. For each colatitude there are two branches. The solid black line delimits the border between attenuation and over-transmission or over-reflexion. The possible cut in the range about $Ro\in[-1.5,0.5]$ corresponds to the stable regime for which the attenuation coefficient is shown in the upper panel.}
\label{astoul:fig234}
\end{figure}

If $\chi<1/4$, the solution near the critical layer is :
\begin{equation}
\Vc(y)=A(y-\yc)^{\frac{1}{2}+\sqrt{\frac{1}{4}-\chi}}+B(y-\yc)^{\frac{1}{2}-\sqrt{\frac{1}{4}-\chi}},
\label{instable}
\end{equation}
where $A$ and $B$ are complex coefficients.
This regime is unstable since waves can be amplified as we will show in the following discussion. The vicinity of the critical layer is decomposed as a three-zone model \citep[see][]{Alvan,LindzenBarker}. Zone II is the unstable region of prescribed length $2\delta$, whereas in the surroundings zone I and III the WKBJ method can be applied. Using the continuity relations between the solutions in these three regions, we are able to determine transmission and reflexion coefficients. We choose the size of zones I and III such that $\chi/(y-\yc)^2$ slowly varies and set its value to $\mK\simeq\chi/\delta^2$. Furthermore, we consider that a wave going from zone I to zone II can be either reflected, or transmitted to zone III. This leads to the following expression for the wave functions in zones I and III :
\begin{equation}
\begin{cases}
\Psi_\mathrm{I}(y)=\e^{i\mK(y-\yc)}+R\e^{-i\mK(y-\yc)} \\
\Psi_\mathrm{III}(y)=T\e^{i\mK(y-\yc)}
\end{cases},
\end{equation}
where $R$ and $T$ are the reflexion and transmission coefficients, respectively . The wave function $\Psi_\I$ is valid in the domain $y-\yc\gtrsim\delta$ while $\Psi_\It$ is valid where $y-\yc\lesssim-\delta$.
In zone II, we use the solution (\ref{instable}).
At both interfaces, the continuity relations for the functions $\Psi$ and their derivatives allow us to determine $A$, $B$, $R$ and $T$. We have plotted the coefficients of reflexion and transmision for a fixed Ekman number $\Ek=10^{-9}$ in the lower panel of Fig. \ref{astoul:fig234}. 
We note that for relatively high Rossby numbers, which correspond to slowly rotating stars at fixed shear or to important differential rotation at fixed global rotation, over-transmission or over-reflexion is possible. Therefore the wave can be attenuated or amplified when going across a critical layer as a function of the rotation (and shear) regime.
\\
\\
To summarize, when $\chi>1/4$, waves are attenuated when going through the critical layer. They transmit their angular momentum to the stable mean flow. Conversely, when $\chi<1/4$, the unstable mean flow provides energy to the waves allowing over-reflexion or over-transmission.

\section{Conclusions}
Our results show how critical layers and interactions between tidal and mean flows are crucial to understand tidal dissipation in differentially rotating stars and planets. The simple analysis that we have carried out can be used to unravel possible regimes that can be observed in direct numerical simulations.
The next step will be to consider the feedbacks of the perturbed wave on the mean flow, to introduce the effects of a magnetic field \citep[e.g.][]{Wei} and to make applications for relevant values of the different dimensionless numbers for stellar and planetary interiors.

\begin{acknowledgements}
A. Astoul, S. Mathis and Q. André acknowledge funding by the European Research Council through the ERC grant SPIRE 647383. The authors ackowledge the PLATO CNES funding at CEA/IRFU/DAp and IRAP.
\end{acknowledgements}

\bibliographystyle{aa.bst}  
\bibliography{astoul.bib}

\end{document}